\documentclass[aps,prl,reprint,10pt,twocolumn,showpacs]{revtex4-1}
 \usepackage{graphicx}
\usepackage{amssymb}
\usepackage{amsmath}
\usepackage{amsfonts}
\usepackage{color}
\usepackage[latin1]{inputenc}
\usepackage{float}
\usepackage{comment}
\usepackage{subfigure}

\newcommand{\lrangle}[1]{\langle{#1}\rangle}

\begin{document}
\title{
Kardar-Parisi-Zhang universality class in $2+1$ dimensions:
Universal geometry-dependent distributions and finite-time corrections}
\author{T. J. Oliveira}
\author{S. G. Alves}
\author{S. C. Ferreira}
\affiliation{Departamento de F\'isica, Universidade Federal de Vi\c cosa,
36570-000, Vi\c cosa, MG, Brazil}

\date{\today}

\begin{abstract}

The dynamical regimes of models belonging to the Kardar-Parisi-Zhang (KPZ)
universality class are investigated in $d=2+1$ by extensive simulations
considering flat and curved geometries. Geometry-dependent universal
distributions, different from their Tracy-Widom counterpart in one-dimension,
were found. Distributions exhibit finite-time corrections hallmarked by a shift
in the mean decaying as $t^{-\beta}$, where $\beta$ is the growth exponent. Our
results support a generalization of the ansatz $h=v_\infty t+(\Gamma
t)^\beta\chi + \eta+\zeta t^{-\beta}$ to higher dimensions, where $v_\infty$,
$\Gamma$, $\zeta$ and $\eta$ are non-universal quantities whereas $\beta$ and
$\chi$ are universal and the last one depends on the surface geometry.
Generalized Gumbel distributions provide very good fits of the distributions in
at least four orders of magnitude around the peak, {which can be used} for
comparisons with experiments. Our numerical results call for analytical approaches
and experimental realizations of KPZ class in two-dimensional systems.
\end{abstract}
\pacs{68.43.Hn, 68.35.Fx, 81.15.Aa, 05.40.-a}

\maketitle


Almost three decades after Kardar, Parisi and Zhang (KPZ)~\cite{KPZ} proposed their
celebrated  equation to describe the coarse-grained regime of evolving surfaces,
a renewed burst of interest on it has been stood out due the experimental
realization of its universality class in turbulent liquid crystal
setup~\cite{TakeSano,*TakeuchiSP,*TakeuchiJSP12} and the achievement
of invaluable analytical
solutions for distinct dynamical regimes and
geometries in $d=1+1$~\cite{SasaSpo1,*Amir,*Calabrese,*Imamura}. The KPZ equation
reads as
 \begin{equation}
 \frac{\partial h (x,t)}{\partial t} = \nu \nabla^{2} h + \frac{\lambda}{2}
(\nabla h)^{2} + \xi,
\label{eqKPZ}
\end{equation}
where $\xi$ is a white noise of mean zero and amplitude $\sqrt{D}$.  
Despite of its original conception for evolving interfaces, the KPZ equation 
has also found its place in others important physical systems~\cite{krugrev}.

A great advance in the theoretical understanding of the KPZ universality
class  has begun at early 2000s with the seminal works of
Johansson~\cite{johansson} and Pr\"ahofer and Spohn~\cite{PraSpo1} presenting
analytical asymptotic solutions of some models in the KPZ class. These solutions
link the height's stochastic fluctuations to universal
distributions~\cite{TW1} of the random matrix theory. Inspired in these exact
results, the ansatz 
\begin{equation}
 h = v_{\infty} t + s_\lambda (\Gamma t)^{\beta} \chi,
\label{eq:ht}
\end{equation}
with the exactly known growth exponent $\beta=1/3$, was conjectured as describing 
the asymptotic interface fluctuations of any
model belonging to KPZ class in $d=1+1$~\cite{PraSpo2,krugrev}. In this
equation, $s_\lambda=\mbox{sgn}(\lambda)$, while the asymptotic velocity $v_{\infty}$ and
$\Gamma$ are model dependent parameters and $\chi$ is a 
{universal random variable with time-independent distribution}
 given by the Gaussian orthogonal ensemble (GOE) for flat
geometries~\cite{johansson,PraSpo1} and the Gaussian unitary ensemble (GUE) for the
curved ones~\cite{PraSpo1,PraSpo2}.
{Notice that, in terms of the constants of the KPZ equation, the parameter $\Gamma$ is given by 
$\Gamma=\frac{1}{2}A^2|\lambda|$ with $A=D/\nu$~\cite{krugrev}.}
These geometry-dependent universality were
confirmed in turbulent crystal liquid
experiments~\cite{TakeSano,TakeuchiSP,TakeuchiJSP12} and in stochastic simulations
of several models without known analytical
solutions~\cite{Alves11,Oliveira12,TakeuchiJstat}.

Many fine-tuning results  have been aggregated to the asymptotic
height distributions (HDs) of one-dimensional KPZ systems. The  limiting processes
describing the surface fluctuations are known as Airy$_1$ and Airy$_2$ processes
for flat~\cite{Sasa1,Boro1} and curved geometries~\cite{PraSpo3}, respectively. 
Finite-time corrections to
Eq.~\eqref{eq:ht} were also analytically~\cite{SasaSpo1,Ferrari}, experimentally~\cite{TakeSano,TakeuchiJSP12},
and numerically~\cite{Alves13} observed, leading to the generalization
\begin{equation}
 h = v_{\infty} t + s_\lambda (\Gamma t)^{\beta} \chi+\eta+\zeta t^{-\beta},
\label{eq:htcorr}
\end{equation}
where $\eta$ and $\zeta$ are non-universal.
The correction $\eta$ introduces a shift in the distribution
of the scaled height
 $q = \frac{h-v_{\infty}t}{s_\lambda(\Gamma t)^{\beta}}$
in relation to the asymptotic distributions. The hallmark of this correction,
a shift in the mean vanishing as $\lrangle{q}-\lrangle{\chi}\sim t^{-1/3}$, has
been verified in the crystal liquid
experiments~\cite{TakeSano,*TakeuchiSP,*TakeuchiJSP12} and computer simulations of
several models~\cite{Alves11,Oliveira12,Alves13}. To our knowledge, only two
exceptions have been reported. In the first one, Ferrari and Frings~\cite{Ferrari} 
analyzed the partially
asymmetric simple exclusion process and found a specific value of the asymmetry
parameter where there is no correction up order $\mathcal{O}(t^{-2/3})$.
Off-lattice simulations of an Eden model consistent with a decay $t^{-2/3}$ were
reported ~\cite{TakeuchiJstat}, but a subsequent analysis showed
that the unusual behavior is an artifact of low precision estimates of
$v_\infty$ and a long crossover to the scaling law $t^{-1/3}$~\cite{Alves13}.  

In contrast to the deep understanding of the KPZ class in $d=1+1$, essentially no exact
results are available in $d=2+1$, the most important dimension for
applications~\cite{barabasi}. Indeed, available analytical
approximations~\cite{Lassig,*Caloiri,*Fogedby} fail in predicting the best
numerical estimates of the scaling exponents~\cite{Kelling}. The best we know about the KPZ class
in $d=2+1$ comes from simulations: The scaling exponents~\cite{Kelling} and
height distributions in the stationary regime~\cite{Marinari,*FabioSS,*Chin} are
accurately known and its universality has been verified. A few works, impaired by
finite-size effects, had studied  height distributions in the dynamical regime
using flat geometry~\cite{landau,FabioGR} when, very recently,
Halpin-Healy~\cite{Healy} reported large-scale simulations
of some KPZ models that convincingly suggest the universality of the height distributions.
Halpin-Healy's analysis is in consonance with our results. 

In the present work, a detailed study of the dynamical regime of several KPZ
models in $2+1$ dimensions is presented. Both flat and curved geometries are
considered. We go beyond the Halpin-Healy's results and show that the
generalized KPZ ansatz given by Eq.~\eqref{eq:htcorr} still holds in $d=2+1$
with the proper growth exponent $\beta=0.24$. The universality of $\chi$, that
differs from the counterparts in 1+1 dimensions, is confirmed and its
geometry-dependence characterized. Also, we have verified that the corrections
in the mean vanish as $t^{-\beta}$ and non-universal  corrections were found for higher
order cumulants. We compensate the absence of an exact analytical expression
for the HDs, showing that generalized Gumbel distributions~\cite{Bramwell}  fit
noticeably well the heights scaled accordingly Eq.~\eqref{eq:htcorr}. 


\textit{Flat geometry} - We performed extensive simulations of three models in
the KPZ class, namely, the restricted solid-on-solid (RSOS)~\cite{KK}, single step
(SS)~\cite{barabasi} and etching~\cite{Mello} models.
Square lattices with up to $2^{15}\times 2^{15}$ sites and
periodic boundary conditions were used. Except for SS model, for which a
checkerboard initial condition was used, an initially smooth substrate was
considered. Up to $10^3$ runs were used in averages.

\begin{figure}[ht]
\includegraphics[width=8cm]{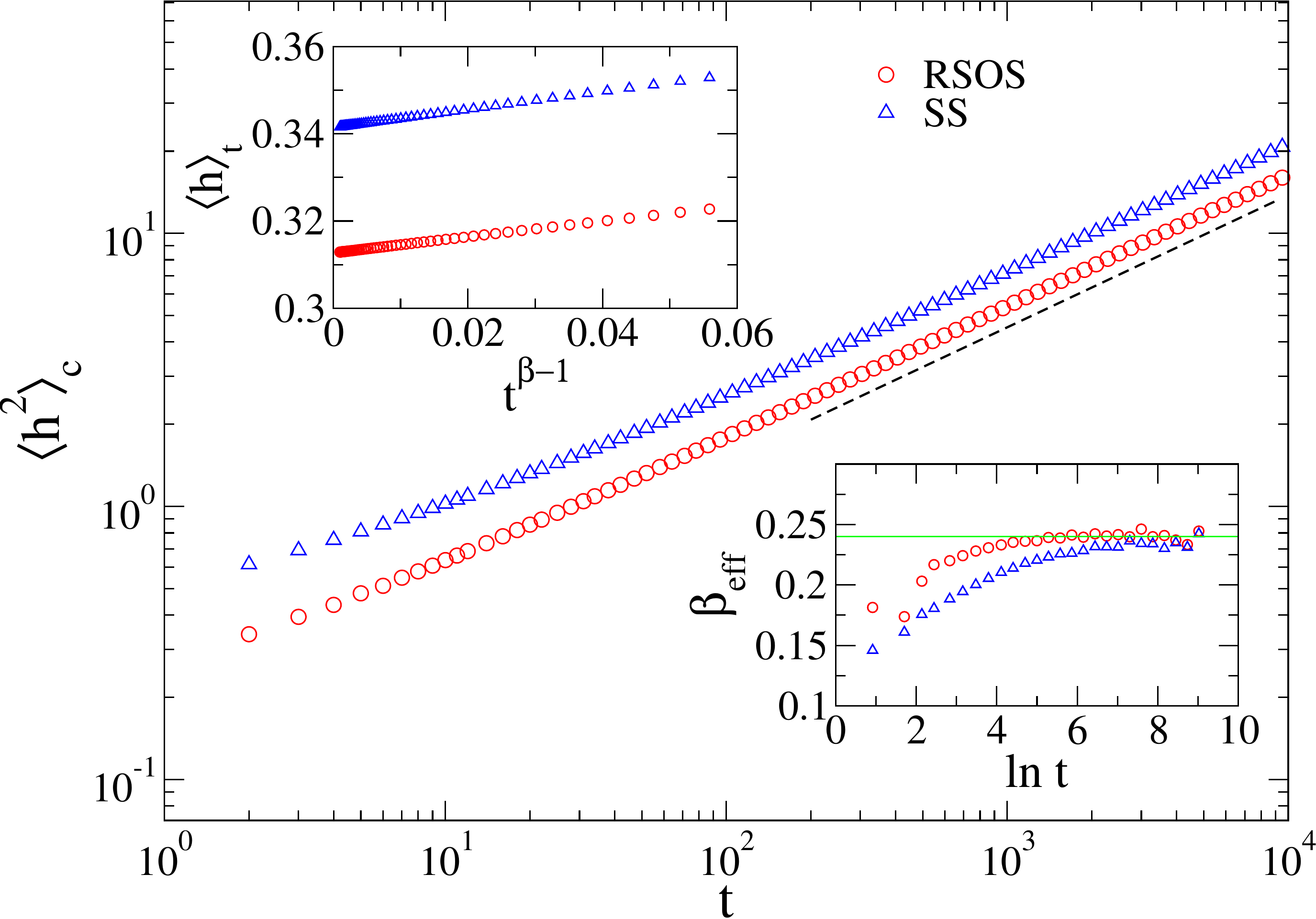}
\caption{(Color online)  Main plot: Height variance against time for RSOS and SS
models. Dashed line represents the power law $t^{0.48}$. Bottom inset:
Effective growth exponents against time. The horizontal line represents the
accepted KPZ value in $d=2+1$. Top inset: Interface velocity against
$t^{\beta-1}$. }
\label{fig:h2cf}
\end{figure}

The growth exponent can be determined from $w^2 \equiv \lrangle{h^2}_c\sim
t^{2\beta}$, where $\lrangle{X^n}_c$ represents the $n$th cumulant of $X$.
Figure~\ref{fig:h2cf} shows the evolution of the variance for two models, while
the corresponding effective growth exponents (local slope in curves $\ln w$
\textit{vs}. $\ln t$) are shown in the bottom inset. The growth exponents
obtained for all models are shown in Tab.~\ref{tab}, in which  an excellent
agreement with the accepted KPZ exponent $\beta=0.24$ is observed for all flat
models. Differentiating $\lrangle{h}$ in Eq.~\eqref{eq:htcorr} one finds 
$\lrangle{h}_t = 
         v_{\infty}+s_\lambda\beta\Gamma^\beta\lrangle{\chi}t^{\beta-1}+\ldots$ 
A linear regression in $\lrangle{h}_t$ against $t^{\beta-1}$ for $t\rightarrow\infty$
yields $v_{\infty}$.
This procedure is illustrated in the top inset of Fig.~\ref{fig:h2cf} 
and the estimates for all investigated models are given in Tab.~\ref{tab}.
 The quantity $\Gamma^\beta \lrangle{\chi}$ can be obtained from the asymptotic value of
$g_1=(\lrangle{h}_t-v_\infty)t^{1-\beta}/\beta$. It was shown that the value of 
$\Gamma$ determined from $g_1$ is more reliable than using cumulants of order $n\ge 2$~\cite{Alves13}.

\begin{figure}[ht]
 \includegraphics[width=7cm]{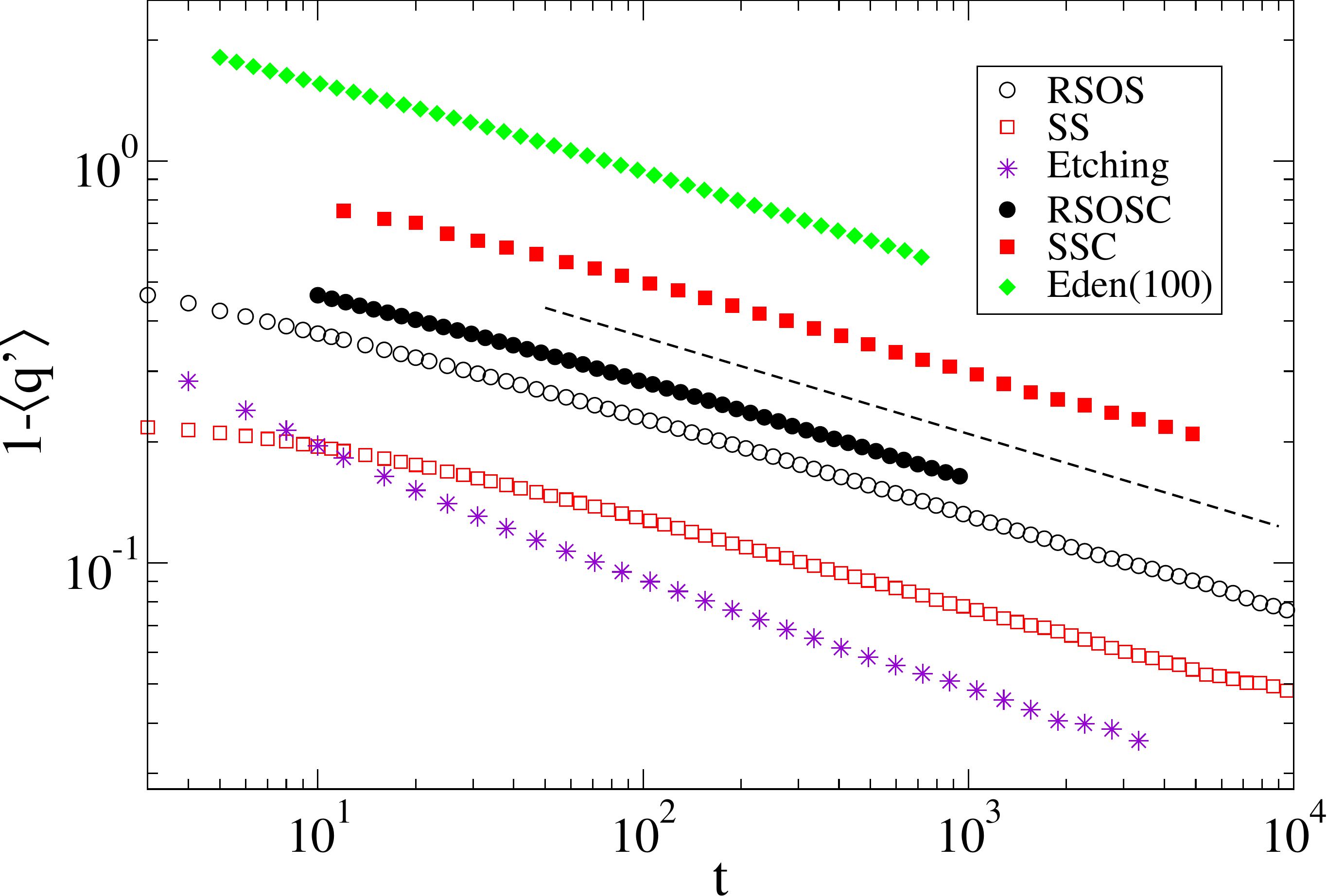}

\includegraphics[width=8cm]{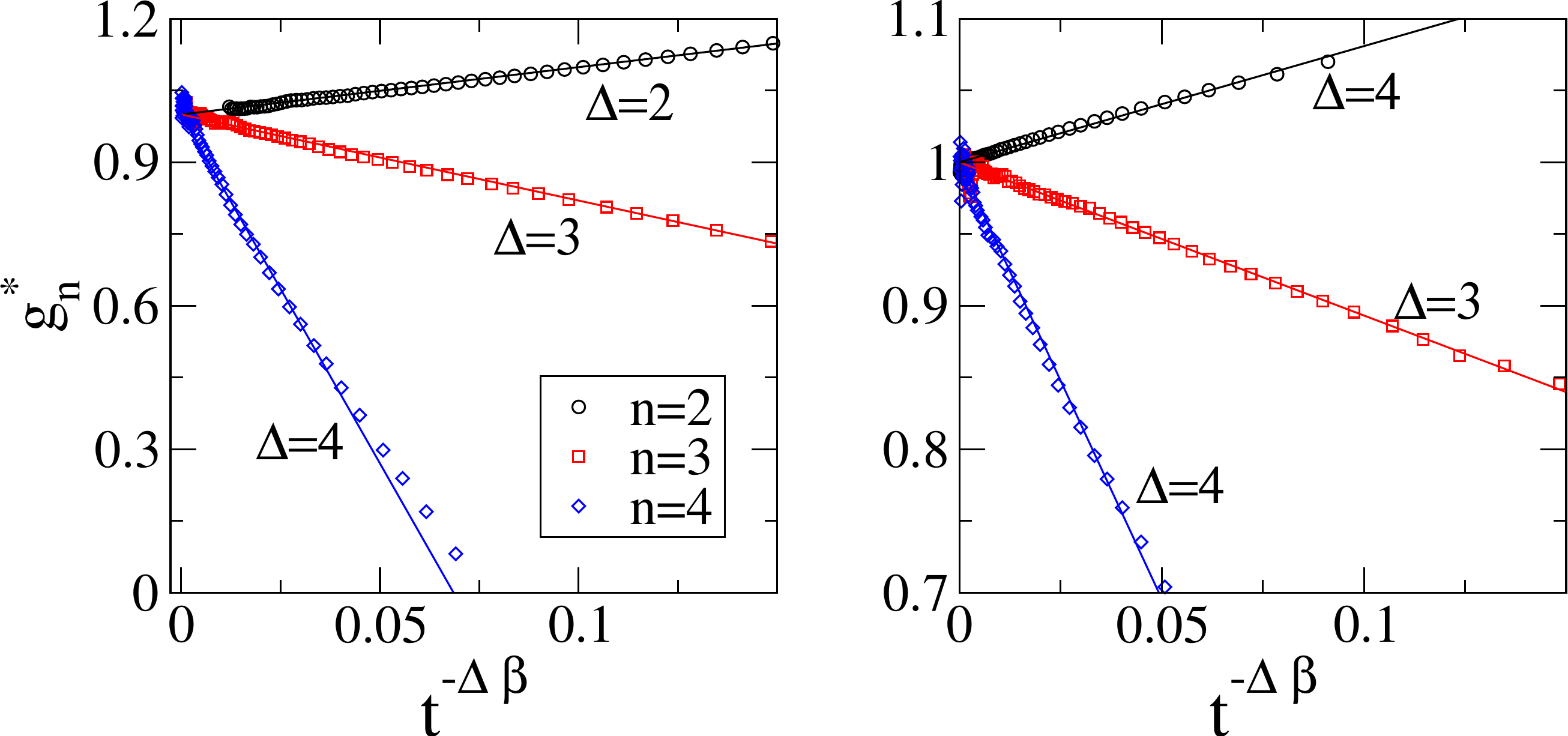}
\caption{Top: Determination of mean shift $\lrangle{\eta}$ for flat and curved
geometries. Dashed line is the decay $t^{-0.24}$. Bottom: Normalized cumulants
$g_n^*=g_n(t)/g_n(\infty)$ against scaled time for SS (left) and RSOS (right)
models in a flat geometry. Lines are (scaled) linear regressions used to determine
$g_n(\infty)$.}
\label{fig:shift1}
\end{figure}

\begin{table*}[ht]
\begin{center}
\begin{small}\begin{tabular}{ccccccccccc}
\hline\hline
model     & $v_{\infty}$&$g_1$      &  $g_2$   &  $g_3$    & $g_4$   &  $\lrangle{\eta}$ & $\beta$ & $R$ & $S$ & $K$\\
\hline
RSOS      &0.31270(1)  &$-0.773(1)$  &0.1936(4) &0.0364(3) &0.0130(5) & $-0.5(1)$ & 0.240(3) 
& 0.324(3)& 0.427(5)  & 0.347(8) \\ 
SS        &0.341368(3) &$-0.881(1)$  &0.250(1)  &0.0536(3) &0.0219(5) & $-0.4(1)$ & 0.239(5) 
& 0.322(2) & 0.428(5)  & 0.35(1)\\ 
Etching   &3.3340(1)   &$-2.348(3)$ &1.715(3)  & 0.950(2)  & 1.00(1)  &0.6(1)   & 0.235(5) 
& 0.311(2) & 0.423(2)   & 0.340(5)\\ 
RSOSC       &$0.3134(2)$& $-2.116(2)$& $0.272(2)$&0.0481(6)& 0.0158(5) &$-1.7(1)$& 0.24(1)  & 0.061(3)    & $0.339(8)$  & 0.21(1) \\
SSC         &  $0.12611(2)$ & $-0.797(1)$& $0.051(2)$ &0.0037(2)  & 0.00053(8) &$-1.2(1)$ & 0.23(2)  & 0.080(5)   & $0.32(4)$   & 0.20(5)\\
Eden(001)   &    $0.6495(3)$   & $-3.543(3)$ & $0.785(8)$ &0.234(3)   & 0.13(1)  & 9.8(5)  & 0.243(7) & 0.063(2) & 0.336(9)  & 0.21(2)\\
Eden(111) &    $0.6242(2)$   & $-3.219(5)$ & $0.610(8)$ &0.164(3)   & 0.083(5) & 8.8(5)  & 0.239(6) & 0.059(2) & 0.34(1)   & 0.22(2)  \\
\hline\hline
\end{tabular}
\caption{\label{tab} Non-universal and universal quantities for the dynamical regime of 
KPZ models. Definitions in the text.}
\end{small}

\end{center}
\end{table*}

The accuracy in determining universality in simulations may be very sensitive to
the correction $\eta$ depending on the model and the attainable simulation time.
So, it is important to determine the strength of corrections before analyzing
height distributions. The mean $\lrangle{\eta}$ 
can be determined using the
height scaled in terms of directly measurable parameters $v_\infty$ and $g_1$ as
$q' = (h-v_\infty t)/(s_\lambda g_1t^\beta)$~\cite{Alves13}. 
Equation~(\ref{eq:htcorr}) implies 
$1-\lrangle{q'}=-(s_\lambda\lrangle{\eta}/g_1)t^{-\beta}+\ldots$
Figure~\ref{fig:shift1} shows that the power law $t^{-\beta}$ describes 
very precisely the shift, analogously as observed in
$d=1+1$~\cite{TakeuchiSP,TakeuchiJSP12,Ferrari,Oliveira12,Alves11,Alves13}.
So, using the prefactor of the power law $t^{-\beta}$, we determined $\lrangle{\eta}$
for all investigated models. The estimates are shown in Tab.~\ref{tab}.

From Eq.~\eqref{eq:htcorr}, we have that scaled cumulants  
$g_{n}(t)=\lrangle{h^n}_c/(s_{\lambda}^{n} t^{n\beta})$, $n \geq 2$, 
converge to $\Gamma^{n \beta} \lrangle{\chi^{n}}_c$ for $t\rightarrow\infty$. 
Contrasting  with the first cumulant, the corrections in $g_n$  depend on the  
model. Figure~\ref{fig:shift1} bottom shows the scaled cumulants 
against $t^{-\Delta \beta}$ where $\Delta$ was assumed integer 
(used values are indicated nearby each curve). For 
sake of visibility, curves were normalized by the asymptotic value 
$\Gamma^{n \beta} \lrangle{\chi^{n}}_c$ obtained by extrapolation in plots
$g_n$ versus $t^{-\Delta\beta}$. These estimates are shown in Tab.~\ref{tab}.
For SS (bottom left in Fig.~\ref{fig:shift1}) and Etching (data not shown) models, the corrections 
are quite consistent with $\lrangle{q^n}_c-\lrangle{\chi^n}_c\sim t^{-n\beta}$, 
in analogy to the exact solution of the KPZ equation with edge initial 
condition and experimental results in $d=1+1$~\cite{SasaSpo1,TakeuchiJSP12}.
However, in RSOS the second cumulant present a 
different behavior with the shift decaying approximately as $t^{-4\beta}$
demonstrating the non-universality of the corrections in cumulants of order $n\ge 2$.

The parameters $g_i$, $i=1$ to $4$, shown in Tab.~\ref{tab} depend on 
$\Gamma$, which can not be determined directly from height distributions~\cite{Alves13}. However,
one can investigate dimensionless cumulant ratios that are independent of $\Gamma$
and, therefore, are expected to be universal.
In Tab. \ref{tab}, we show the ratios 
$R=g_2/g_1^2 = \lrangle{\chi^2}_c/\lrangle{\chi}^2$, 
$S=g_3/g_2^{3/2}=\lrangle{\chi^3}_c/\lrangle{\chi^2}_c^{3/2}$ (skewness) and 
$K=g_4/g_2^2=\lrangle{\chi^4}_c/\lrangle{\chi^2}_c^{2}$ (kurtosis) for all  
investigated models. The ratios for different flat models are
essentially the same, confirming the universality of $\chi$ conjectured initially. 
Notice that they are different from the ratios for GOE distributions 
expected for their one-dimensional counterparts~\cite{PraSpo1}.
Since an infinite  hierarchy of cumulant ratios can be measured, in principle,
we can determine all cumulants in terms of the first one. 
Our estimates for $S$ and $K$ are in 
good agreement with those found by Halpin-Healy in~\cite{Healy}, but fluctuating estimates for
$\lrangle{\chi}$ and $\lrangle{\chi^2}_c$ presented there do not allow a 
reliable estimate of $R$ (values ranging from 0.33 to 0.51 are extracted from
Ref.~\cite{Healy}). We believe that the corrections in distributions, 
mainly in the mean, are responsible by the apparent non-universality 
of $R$ in Ref.~\cite{Healy}. Our estimates  of $S$ and $K$ are also consistent 
with former, small-size simulations~\cite{landau}
and also with recent simulations of Eden model on flat substrates~\cite{Alves12}, confirming the universality of the HDs.

Due to the lack of rigorous results in 2+1 dimensions, we are currently 
not able to associate our numerical results to an analogous of 
TW distributions. However, previous works {dealing with linear systems}
have shown that the generalized  
Gumbel distribution with a non-integer parameter $m$ 
fits the probability density functions
of stationary quantities in several equilibrium and non-equilibrium 
systems~\cite{Bramwell,Antal,Lee}. 
We have obtained a very good agreement between our simulations and 
the so-called Gumbel's first asymptotic distribution of mean $\lrangle{X}$
and variance $\lrangle{X^2}_c$~\cite{Lee},
\begin{equation}
G(X;m)=\frac{m^mb}{\Gamma(m)} \exp\left[-m\left(z_X+e^{-z_X}\right)\right],  
\label{eq:gumbel}
\end{equation}
where $b=\sqrt{\psi_1(m)/\lrangle{X^2}_c}$, $z_\chi=b(\lrangle{X}-X+s)$,
 $s=[\ln m-\psi_0(m)]/b$, $\Gamma(X)$ is the gamma 
function and $\psi_k(X)$ the polygamma function of order $k$~\cite{Gradshtein}. 
The parameter $m$ allows to change simultaneously, but not independently, 
the skewness and kurtosis of the distribution. For $m=6$, one obtains 
a skewness $S_G=-0.4247$ and kurtosis $K_G = 0.3597$ very close to the universal
values for flat models shown in Tab.~\ref{tab}.

\begin{figure}[!t]
\begin{center}
 \includegraphics[width=7.5cm]{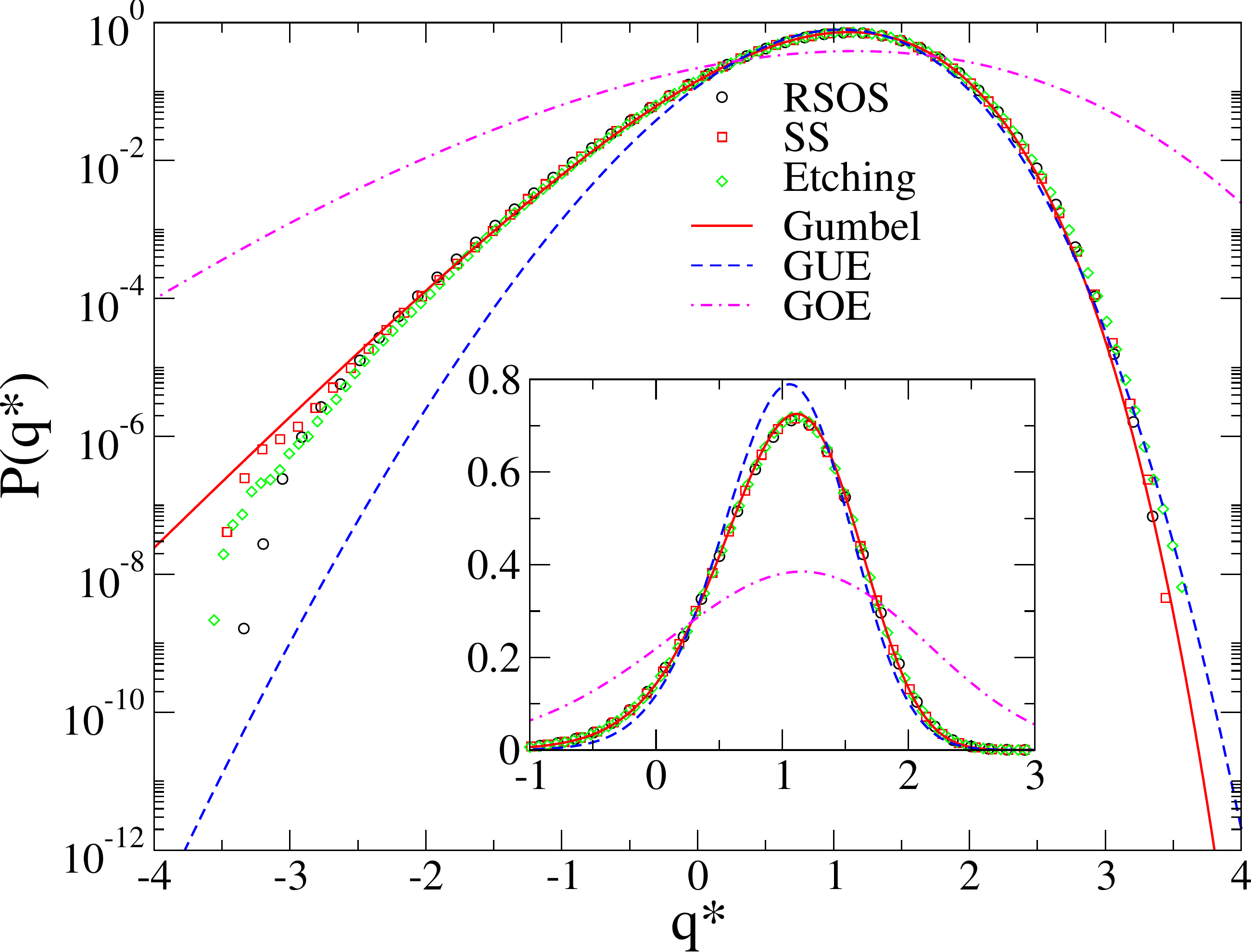}
\end{center}
\caption{(Color online)
Height distributions for {flat growth} scaled to mean 1 compared
with a Gumbel distribution for $m=6$ and variance $R=0.32$.  The
inset shows the same data in a linear scale. The growth times 
are: $t=10^4$ (RSOS), $t=8000$  (SS), and $t=2000$ (etching).
Scaled TW distributions are included for sake of comparison. 
}
\label{fig:pofn}
\end{figure}

The height distribution scaled to a mean 1, accordingly the non-universal parameters,  
becomes
$q^* = ({h-v_\infty t - \lrangle{\eta}})/({s_\lambda g_1t^\beta}), $
leading to a variance $\lrangle{{q^*}^2}_c\equiv R$.
In top panel of Fig. \ref{fig:pofn}, the scaled heights for flat models are
compared with a Gumbel distribution for $m=6$, mean 1  and variance
$R=0.32$. A remarkable collapse is observed around the peak for at least four 
decades.
From an experimental
perspective, it is extremely hard to measure distribution extremes with an
accuracy comparable to our simulations. Hence, the Gumbel approximation {
is  a useful reference} to check the KPZ universality class in 2+1 dimensions. Notice
that in a linear scale, simulations are indistinguishable from the Gumbel
distribution in contrast with the TW distributions that not even barely fit the
distribution's peak as can be seen in inset of Fig.~\ref{fig:pofn}. 
Interestingly, the rightmost tail of the scaled 
distributions is well fitted by the scaled GUE distribution 
$\chi_{gue}/\lrangle{\chi_{gue}}$. 
It is worth mentioning that generalized Gumbel functions
was compared with distributions of height extremes in the stationary regime
of  KPZ and other non-linear models in Ref.~\cite{tiago2}. A 
good fit around the peak and large deviations in the tails were observed.

\textit{Curved geometry -} We study radial geometry using the on-lattice Eden~D
model~\cite{Alves13}. Due to the intrinsic anisotropy of on-lattice Eden
clusters, we investigate surface fluctuations along axial (100) and diagonal
(111) directions. We also considered curved surfaces using  the RSOS and SS
models growing in a corner (RSOSC and SSC), where fluctuations in (111)
direction are considered. Details of the models and simulation 
are presented in Ref.~\cite{Alves13}, where we carried out a detailed
study in $d=1+1$ and obtained the expected KPZ scaling, GUE
TW, for curved growth. 

\begin{figure}[!t]
\begin{center}
 \includegraphics[width=8cm]{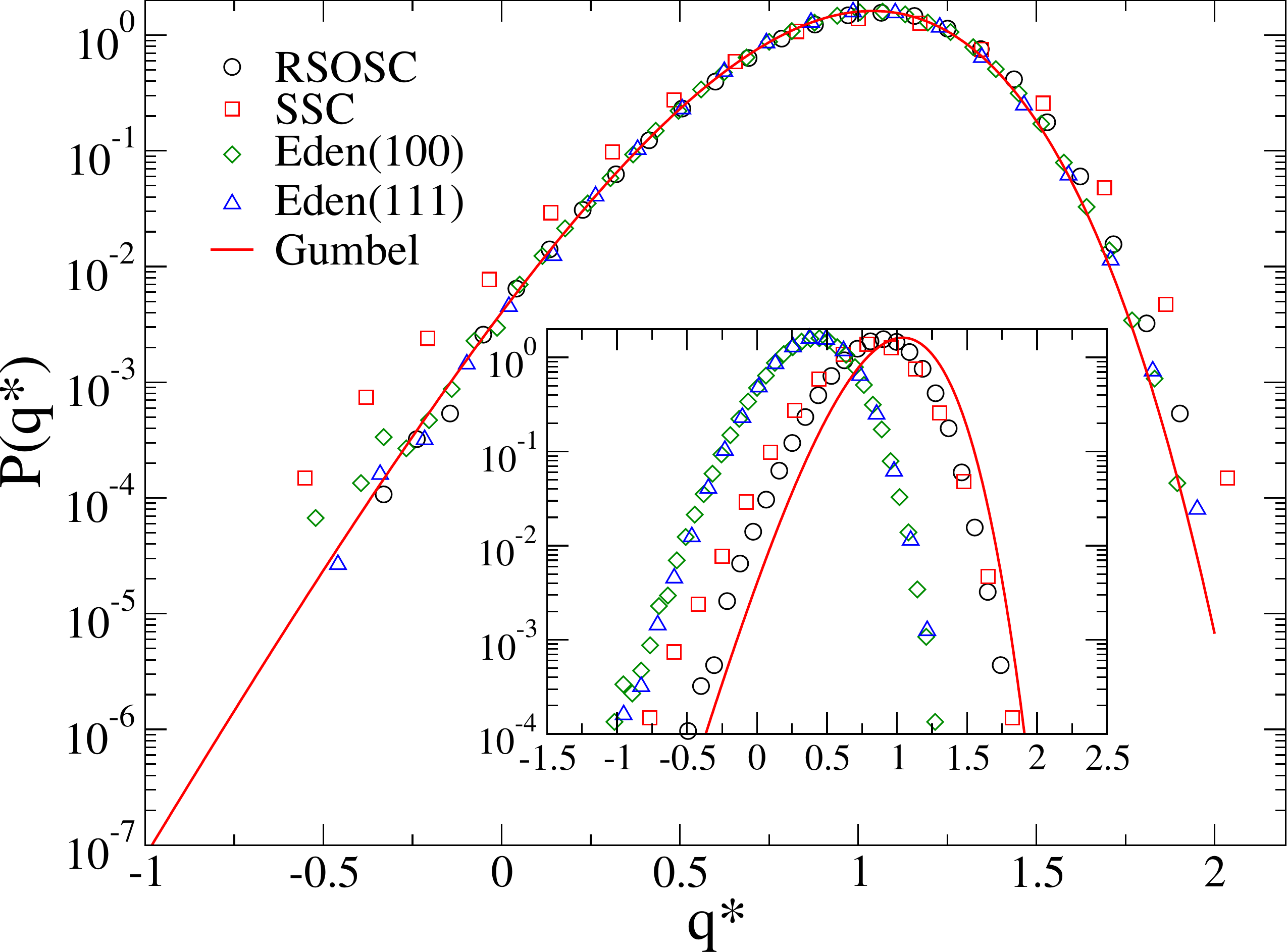} 
\end{center}
\caption{(Color online) Height distributions {for curved growth} scaled to mean 1 compared
with a Gumbel distribution for $m=9.5$ and variance $R=0.062$.
The growth times are : $t=1012.4$ (RSOSC), $t=4000$ (SSC), and $t=549.7$ (Eden).
Inset: Scaled height distributions disregarding the shift $\lrangle{\eta}$.}
\label{fig:pofhcurv}
\end{figure} 

The growth exponents found for all models agree very well with the KPZ value
$\beta=0.24$, as shown in Tab.~\ref{tab}. The non-universal parameters related
to are shown in Tab.~\ref{tab}. The asymptotic velocity of SSC model has been
under debate~\cite{Olejarzreply,Rajeev} and our estimate is in agreement with
Ref.~\cite{Rajeev}. Again, the shift in the mean scales as $t^{-\beta}$ exactly
as in the flat case (Fig.~\ref{fig:shift1}). However, the amplitude of the
corrections are in general much larger than in the flat case, particularly for
Eden model, and plays a central role for the time scale simulated in the present
work. Corrections in $g_n$ are of order $t^{-2\beta}$ or faster in
analogy to the flat case.

Dimensionless cumulant ratios are also universal for curved geometries as shown
in Tab.~\ref{tab}. These ratios differ from those of the flat case and are even
further from the TW values known for 1+1 dimensions. Our cumulant ratios are
also in agreement with those reported by Halpin-Healy for a single model in the
so-called point-point geometry~\cite{Healy}. Once again, the scaled height
distributions are well fitted by a generalized Gumbel distribution with $m=9.5$,
which has $S_G=0.335$ and $K_G=0.224$. A very important remark is that curves do
not collapse if the correction $\lrangle{\eta}$ is not explicitly included in
the analysis as shown in the inset of Fig.~\ref{fig:pofhcurv}. Rescaling the
distributions, accordingly to Eq.~(\ref{eq:htcorr}), to mean $1$ and variance $R =
0.062$ we found a good data collapse, with exception of the SS model
(Fig.~\ref{fig:pofhcurv}). This is due to its larger value of $R$
(possible produced by large fluctuations). 

Assuming the last term in Eq.~(\ref{eq:htcorr}) has the form $\zeta t^{-\gamma}$, 
one has that 
\begin{equation}
s_\lambda(\lrangle{h}_t-v_\infty)t^{1-\beta} =  g_1
-\gamma s_\lambda\lrangle{\zeta} t^{-\gamma-\beta}.
\end{equation}
Our simulations show that $g_1$ converges to its asymptotic value with a
correction quite close to $t^{-2\beta}$ in all flat and curved growth models. 
So, the last term in Eq.~(\ref{eq:htcorr}) decays with an
exponent $\gamma=\beta$. An equivalent result was obtained in the simulations of
KPZ models in $d=1+1$ where a term $t^{-1/3}$ was identified in the KPZ 
ansatz~\cite{Alves13}. So, we have an additional evidence that 
the generalized KPZ ansatz in $d=1+1$ has an equivalent counterpart in higher
dimensions.

In conclusion, we have studied the height distributions in the dynamical regime
of KPZ systems in $d=2+1$ and confirmed the universality of geometry-dependent
distributions found very recently by Halpin-Healy~\cite{Healy}. However, we
have gone further and characterized also the finite-time behavior of the
distributions. As in the $1+1$ case, the shift in the mean decays as
$t^{-\beta}$ and the corrections in higher order cumulants are non-universal and
decay faster or equal than $t^{-2 \beta}$. We also show that generalized Gumbel
distributions, commonly applied to fit distributions in linear systems~\cite{Bramwell,Antal,Lee},  
fit noticeably well the HDs of KPZ models that are non-linear. 
Such distributions and the finite-time behaviors
{may play an import rule in the} experimental study of KPZ systems. Furthermore,
they {may} motivate and guide analytical insights to the understanding of the KPZ
universality class in two dimensions.

\begin{acknowledgments}
Authors acknowledge the support from CNPq and FAPEMIG (Brazilian agencies).
\end{acknowledgments}


%

\end{document}